\documentclass{vgtc}                          
\ifpdf
  \pdfoutput=1\relax                   
  \usepackage{graphicx}                
  \usepackage{subcaption}
  \DeclareGraphicsExtensions{.pdf,.png,.jpg,.jpeg} 
\else
  \usepackage{graphicx}                
  \usepackage{subcaption}
  \DeclareGraphicsExtensions{.eps}     
\fi%

\graphicspath{{figures/}{pictures/}{images/}{./}} 

\usepackage{microtype}                 
\PassOptionsToPackage{warn}{textcomp}  
\usepackage{textcomp}                  
\usepackage{mathptmx}                  
\usepackage{times}                     
\usepackage{enumitem}
\usepackage{cite}                      
\usepackage{tabu}                      
\usepackage{booktabs}                  
\usepackage{xspace}
\usepackage{url}

\onlineid{0}

\vgtccategory{Research}

\newcommand{\systemname}{DBenVis\xspace}

\newcommand{\eg}{{\em e.g.}}

\title{DBenVis: A Visual Analytics System for Comparing DBMS Performance via Benchmark Programs}

\author{Yoojin Choi \\
        \scriptsize Seoul National University \\
        \scriptsize Seoul, Korea \\
        \scriptsize cyj@dbs.snu.ac.kr
\and Juhee Han\\
     \scriptsize Seoul National University \\
     \scriptsize Seoul, Korea \\
     \scriptsize juheehan@dbs.snu.ac.kr
\and Daehyun Kim\\
    \scriptsize Seoul National University \\
    \scriptsize Seoul, Korea \\
    \scriptsize creative26@dbs.snu.ac.kr}

\teaser{
  \centering
  \includegraphics[width=\linewidth]{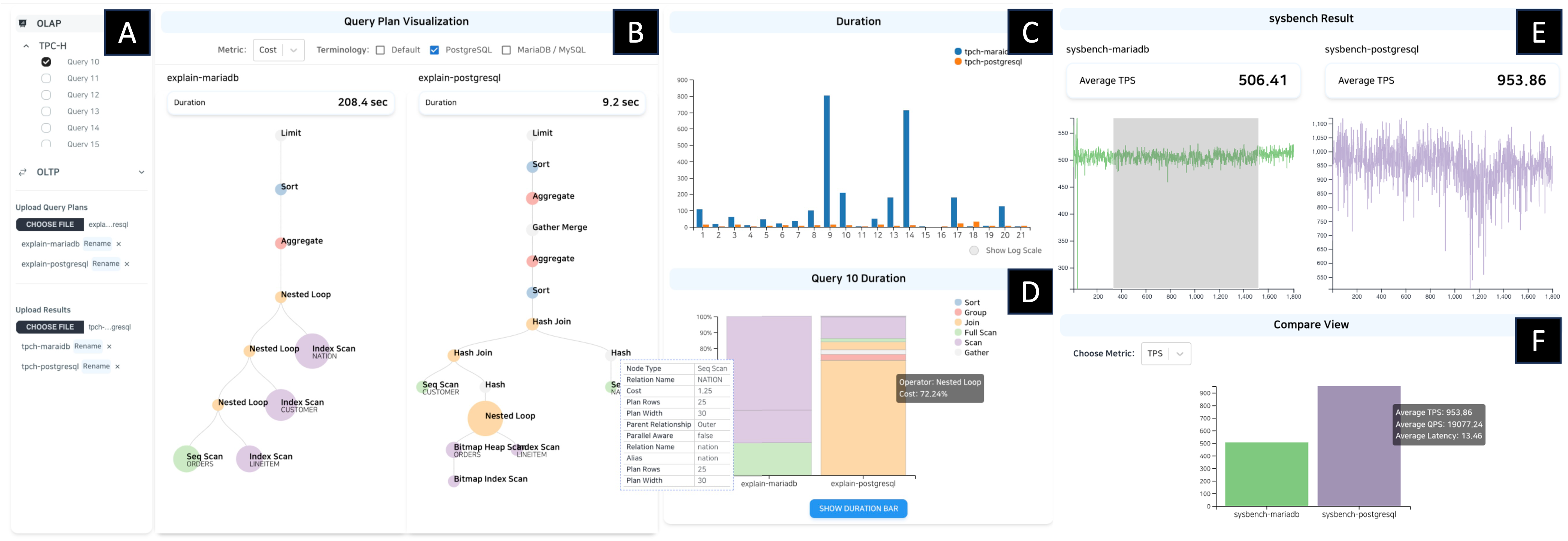}
  \label{fig:overview}
  \caption{\systemname system.
  \textbf{A.} Users can choose benchmarks and upload, delete, or rename their result files.
  \textbf{B.} Query Plan Visualization shows query execution plans as collapsible trees, where the size of each node is proportional to a selected metric.
  \textbf{C.} TPC-H Duration provides a comparison of the performance between DBMSs.
  \textbf{D.} The TPC-H Specific Query Duration provides a detailed performance analysis of the selected query.
  \textbf{E.} The sysbench Result visualizes time series data of selected metrics.
  \textbf{F.} The Compare View supports the comparison of average values for the selected metrics.}
}

\abstract{Database benchmarking is an essential method for evaluating and comparing the performance characteristics of a database management system (DBMS).
It helps researchers and developers to evaluate the efficacy of their optimizations or newly developed DBMS solutions.
Also, companies can benefit by analyzing the performance of DBMS under specific workloads and leveraging the result to select the most suitable system for their needs.
The proper interpretation of raw benchmark results requires effective visualization, which helps users gain meaningful insights.
However, visualization of the results requires prior knowledge, and existing approaches often involve time-consuming manual tasks.
This is due to the absence of a unified visual analytics system for benchmark results across diverse DBMSs.
To address these challenges, we present \systemname, an interactive visual analytics system that provides efficient and versatile benchmark results visualization.
\systemname is designed to support both online transaction processing (OLTP) and online analytic processing (OLAP) benchmarks.
\systemname provides an interactive comparison view, which enables users to perform an in-depth analysis of performance characteristics across various metrics among different DBMSs.
Notably, we devise an interactive visual encoding idiom for the OLAP benchmark to represent a query execution plan as a tree.
In the process of building a system, we propose novel techniques for parsing meaningful data from raw benchmark results and converting the query plan to a D3 hierarchical format.
Through case studies conducted with domain experts, we demonstrate the efficacy and usability of \systemname.}

\begin{document}

\firstsection{Introduction}
\maketitle

In the field of database management systems, performance analysis is one of the most important aspects of data management\cite{oltp-bench}.
Hence, database benchmarking is essential for evaluating and comparing the performance characteristics of diverse DBMSs across varying workloads.
DBMS researchers and developers benefit from database benchmarking, using it to compare multiple DBMSs in terms of performance, query plan, and resource utilization.
It also helps evaluate the performance of their optimization or newly-developed DBMS compared to existing ones.
Companies can also benefit by gaining valuable insights from the results and making an informed decision when selecting the most suitable DBMS for their needs.
In order to fully leverage the benchmarking tools, such as in the cases described above, adequate visualization of benchmark results is necessary.

However, the process of visualizing the benchmark results is often cumbersome, involving the extraction of data, selection of suitable visualization tools, and the actual execution of visualization.
This time-consuming procedure requires specialized expertise in benchmarks and visualization tools.
Furthermore, there lacks a unified system that allows users to automatically compare benchmark results across multiple DBMSs.
Another major challenge is the visualization of query execution plans, which are crucial for performance analysis.
While existing tools such as PostgreSQL Explain Visualizer~\cite{pev} and MySQL Visual Explain Plan~\cite{mysql} provide visualization of query plans, the problem is that they are limited to specific DBMS and use distinct visualization designs.
This results in the absence of a comprehensive, cross-DBMS visual analytics solution.

To address these challenges, we introduce \systemname, an interactive visual analytics system designed to provide user-friendly and effective visualizations of benchmark results and query plans.
\systemname supports both OLTP and OLAP workload and provides handy interaction features.
The system employs a line chart to depict time series data of various performance metrics and a bar chart to enable comparative analysis for OLTP benchmarks.
For OLAP benchmarks, the system uses various bar chart types, such as grouped and stacked bar charts, for a thorough performance analysis. It also visualizes query plans as trees.
We also evaluate our system through case studies with domain experts to confirm the system's efficacy and usability.

The major contributions of this paper are as follows:
\begin{itemize}
    \item We develop techniques for extracting essential data from raw benchmark results and parsing and rearranging \verb|EXPLAIN| results of various DBMSs into a D3 hierarchical format.
    \item We implement novel and unified visualization designs tailored to OLAP and OLTP benchmark results and query plans.
    \item We provide a user-friendly application designed to visualize diverse metrics comprehensively, providing detailed analysis.
\end{itemize}

The rest of the paper is organized as follows. Section~\ref{sec:related} covers related works, and Section~\ref{sec:req} defines requirements and tasks.
Section~\ref{sec:dbenvis} presents a comprehensive explanation of \systemname and its contributions.
In Section~\ref{sec:eval}, we present the evaluation results.
Lastly, the conclusion with discussions and future works is covered in Section~\ref{sec:conclusion}.

\section{Related Work} 
\label{sec:related}
We introduce the research areas related to \systemname, primarily focusing on two domains: benchmark result visualization and query plan visualization.
Our work gets inspiration from existing studies and attempts to overcome some of the limitations in these areas.

\subsection{Benchmark Result Visualization}
Benchmark result visualization is a key aspect in the analysis of database performance.
Several popular libraries are used in the visualization process, each with unique features and requirements.
Matplotlib~\cite{Hunter:2007} is among the most popular Python libraries aimed at creating interactive visualizations.
D3~\cite{Michael:2011} is one of the most famous web-based JavaScript libraries for visualization.
Numerous other visualization libraries exist, including Seaborn~\cite{Waskom2021} and FusionCharts. 
Although these libraries serve as a powerful tool for visualization, the challenge is that they require users to preprocess the raw benchmark results.
This requires users to have prior knowledge of result data and the library, making it time-consuming. 

HammerDB~\cite{hammerdb} is an open-source database benchmarking tool that allows loading and evaluating DBMS performance using TPC benchmarks and facilitates result visualization.
However, its utility is constrained since it supports only a limited range of benchmarks and database systems.
These limitations highlight the need for more versatile and user-friendly tools in database benchmark visualization.

\subsection{Query Plan Visualization}
The query planner in DBMS generates possible plans for submitted queries and examines each of these possible execution plans, ultimately selecting the execution plan that is expected to run the fastest.
The selection of the right plan to match the query structure and the properties of the data is critical for good performance, so the system includes a complex planner that tries to choose the good plan.
\verb|EXPLAIN| command is used to see what query plan is chosen.
The query plan generated from \verb|EXPLAIN| structure follows a tree format, with scan nodes at the bottom level of the tree.
For queries that require joining, aggregation, sorting, or other operations, additional nodes are present to execute these operations.

There are only a few query plan visualizers available.
The PostgreSQL Explain Visualizer (PEV)~\cite{pev} provides an interactive interface for visualizing query plans in a tree format.
Users can engage with the visualization through clicking, zooming, and panning interactions.
Although not as comprehensive as PEV, MySQL does provide its own visualization tool for explaining plans.
This tool facilitates a basic understanding of query execution plans within the MySQL environment.
Additionally, as far as we surveyed, no query plan visualizer is available for MariaDB.
This notable absence highlights the need for a unified visualizer to compare and analyze query plans across various DBMSs.

\section{Requirements and Tasks}
\label{sec:req}
We conducted a survey to identify limitations in existing works related to benchmark results and query plan visualization.
Based on the challenges defined, we organized them into three analytics tasks and formulated corresponding requirements.

\subsection{Task abstraction}
We formulated three significant tasks and corresponding questions for each task as follows:

\begin{itemize}
    \item[T1.] \textbf{Comprehensive DBMS evaluation.} This task involves an evaluation of DBMS performance across various benchmark programs, including resource usage and query plans. \textit{What are the average duration or TPS for each DBMS? }
    \item[T2.] \textbf{In-depth DBMS comparison.} This task aims to compare DBMSs extensively by considering various metrics (\eg, performance and query plan). \textit{Which operation acts as a bottleneck? What are the major differences between two DBMSs in terms of performance? How do they vary in handling rows and costs?}
    \item[T3.] \textbf{Explore and analyze details.} This task contains exploring additional details (\eg, expected cost of an operator in query plans). \textit{ What are the cost implications of specific query plan operators?}
\end{itemize}

\subsection{Requirement Analysis}
Based on the above tasks, we formulated the requirements that \systemname must satisfy as follows:

\begin{itemize}
    \item[R1.] \textbf{Visualize benchmark results.} The system should provide a comprehensive view of DBMS performance on benchmark programs. OLTP benchmarks include visualizing the TPS in time series and the average TPS. For OLAP benchmarks, it should visualize the duration of each query and query plans during the query execution. (T1 and T2)
    \item[R2.] \textbf{Support comparison.} The system should support comparisons between DBMSs regarding performance and query plans. Users should be able to add or delete DBMSs, and the view should respond interactively. (T2)
    \item[R3.] \textbf{Provide interactive dashboard.} The system should provide interactive features to support in-depth analysis. Users can upload result files and define the DBMS name or use predefined names as legends in plots. For the OLTP workload, users can select the specific range of time they want to explore. For OLAP, users can select the specific queries they want to explore. (T3)
\end{itemize}

\section{The DBenVis System}
\label{sec:dbenvis}
We introduce \systemname, a visual analytics system for DBMS researchers and developers seeking to visualize benchmark results in an easy and efficient way. This section describes data reconstruction techniques developed to extract meaningful data from benchmark results and rearrange query plans into hierarchical format. Also, novel visualization designs within \systemname are presented.

\subsection{Interactive File Managements}
\begin{figure}[t]
    \centering
    \includegraphics[width=0.9\linewidth]{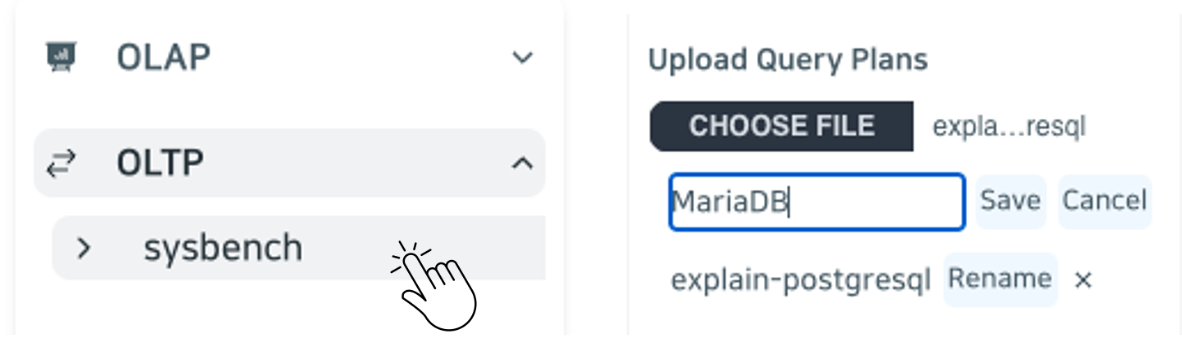}
    \caption{User interface for benchmark selection and file management.}
    \label{fig:file-input}
\end{figure}

This section introduces the interactive file management features in \systemname.
The process involves selecting a benchmark, uploading raw result files, and interactively modifying the view.
\systemname provides users with the option to choose between benchmarks. It currently supports two benchmarks: sysbench\cite{sysbench} and TPC-H\cite{tpch}.
Users can proceed to upload the corresponding result files obtained from benchmark execution.
Multiple files can be uploaded for comparison purposes, with the interface adjusting responsively.
Also, the interface allows the deletion of unwanted files and the renaming of files to apply customized names in the charts.
Figure~\ref{fig:file-input} illustrates the benchmark selection and file upload view of \systemname.

\subsection{Online Transaction Processing Benchmark}
\begin{figure}[t]
    \centering
    \includegraphics[width=0.9\linewidth]{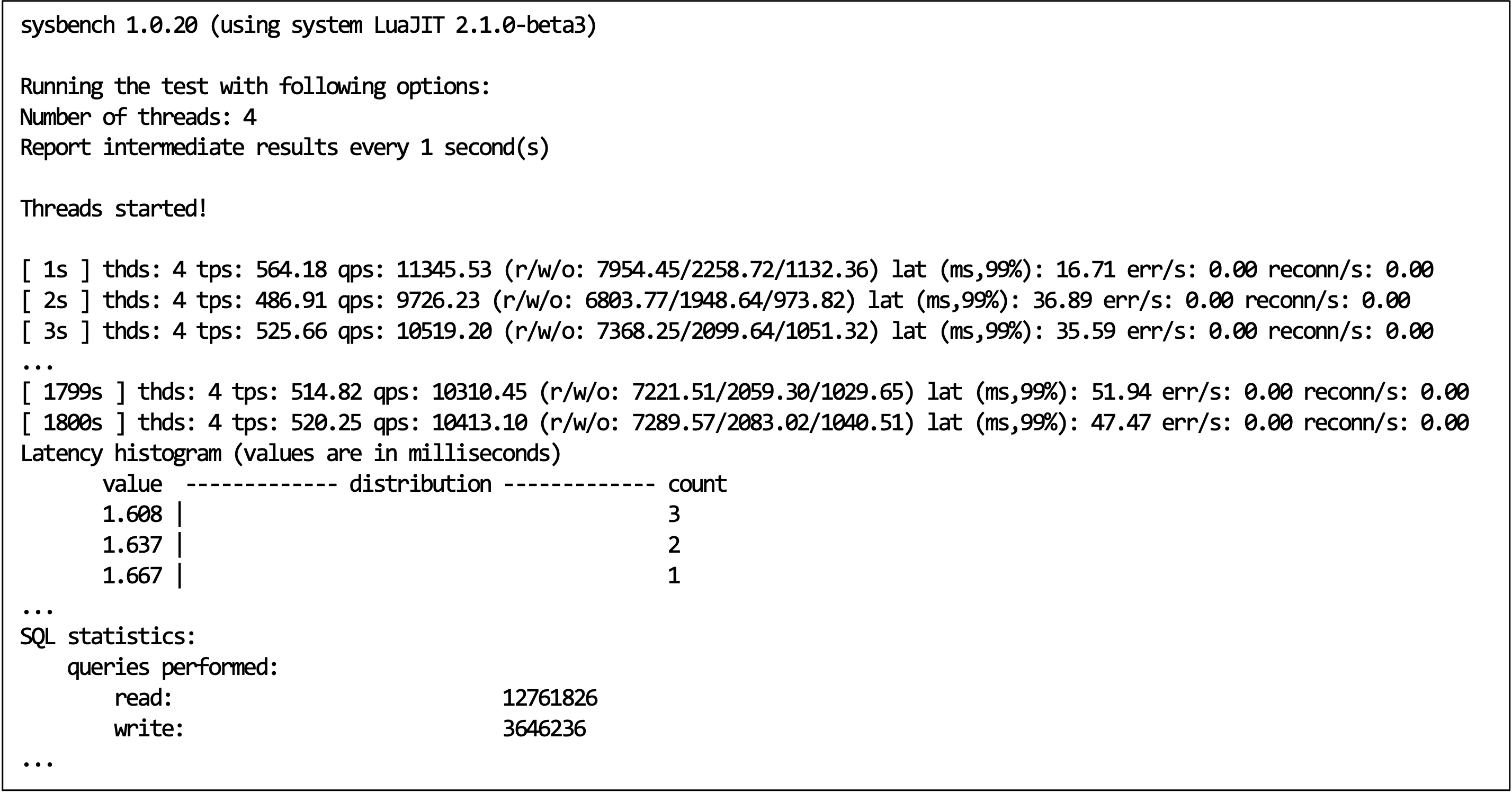}
    \caption{Partial snapshot of the result obtained from sysbench execution.}
    \label{fig:raw-sysbench}
\end{figure}

Our system supports sysbench, an open-source multi-threaded benchmarking tool designed to execute OLTP workloads.
Figure~\ref{fig:raw-sysbench} illustrates a partial snapshot of the result obtained from sysbench execution. It outputs a snapshot of workload statistics every second, including metrics such as transactions per second (TPS), queries per second (QPS), and latency per transaction in milliseconds. 
The final report provides users with average metrics values.

One challenge arises from the extensive set of intermediate results presented on a second-by-second basis, which makes it difficult for users to analyze the overall performance.
The problem worsens when there are multiple result files to compare.
Also, the final report lacks flexibility, as users cannot specify the time range for the average calculation.
This can be a critical issue when the time range that exhibits unstable performance (\eg, early phase of benchmark execution) should be excluded from the average calculation.

\subsubsection{Data Processing for sysbench Results}
We outline the data preprocessing techniques implemented to address the challenges presented above.
Our approach involves parsing the raw result file, organizing each statistical metric data into an array format arranged chronologically, and subsequently passing the processed data to the module responsible for chart generation.
The detailed process proceeds as follows: Upon the user submission of a file, \systemname parses the results and stores TPS, QPS, and latency.
Additionally, the system calculates the average values for each metric.
When a user brushing event is dispatched, the system recalculates values exclusively based on the selected data points.
The updated averages are then applied to visualization components.

\subsubsection{sysbench Result Visualization Designs}
\begin{figure}[t]
    \centering
    \includegraphics[width=0.9\linewidth]{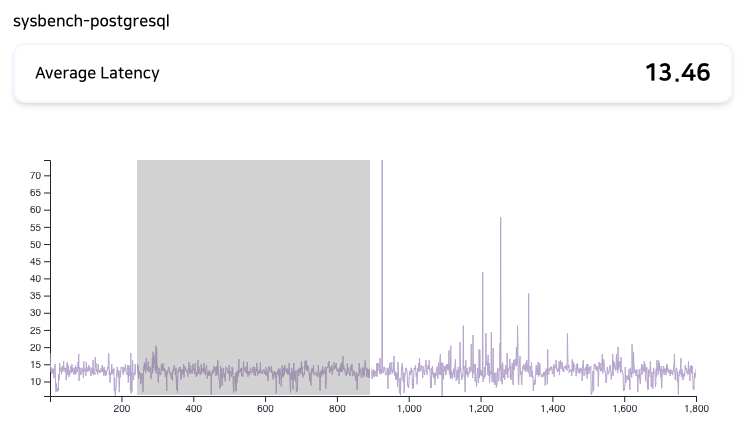}
    \caption{The card component and line chart with brush interaction for detailed analysis.}
    \label{fig:brush}
\end{figure}

\begin{figure}[t]
    \centering
    \includegraphics[width=0.9\linewidth]{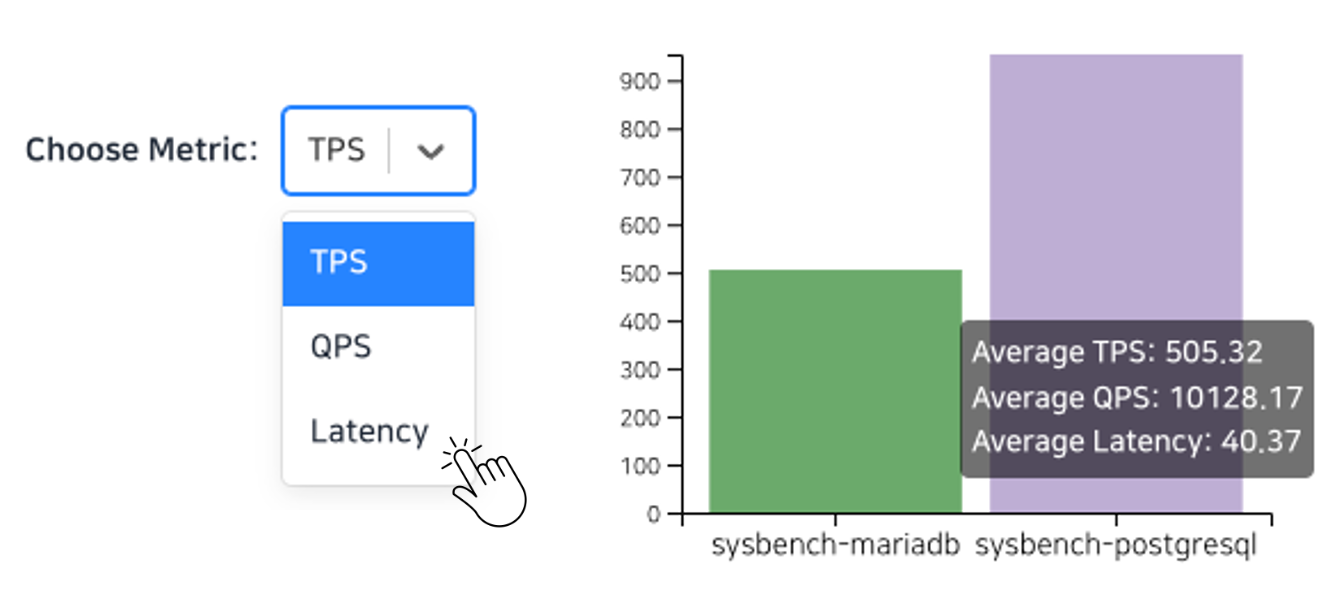}
    \caption{Interaction features of metric selection and tooltips.}
    \label{fig:sysbench-interact}
\end{figure}

This section presents the design of visualizations and interaction features aimed to ensure an effective and user-friendly interface.
The processed metric data are visualized as a line chart, an effective chart type for visualizing trends of time series data over time intervals.
Each result file is visualized through an individual line chart, with a corresponding card component above, which provides the average value of the selected metric.
The bar chart positioned below enables an intuitive comparison of average values.
We employ a bar chart because of its excellence in depicting the relationship between numeric values.

As shown in Figure~\ref{fig:brush}, the line chart contains brush and zoom features, allowing users to focus on desired time ranges.
When users incur a brushing event, the chart zooms in on the selected range, and average values are dynamically updated in the card component and bar chart.
Figure ~\ref{fig:sysbench-interact} demonstrates how users can select a specific metric to visualize, which provokes responsive adjustments in the line chart and bar chart.
Also, tooltips are implemented in the bar chart to facilitate users viewing all three metrics at once.

\subsection{Online Analytic Processing Benchmark}
\begin{figure}[t]
    \centering
    \includegraphics[width=\linewidth]{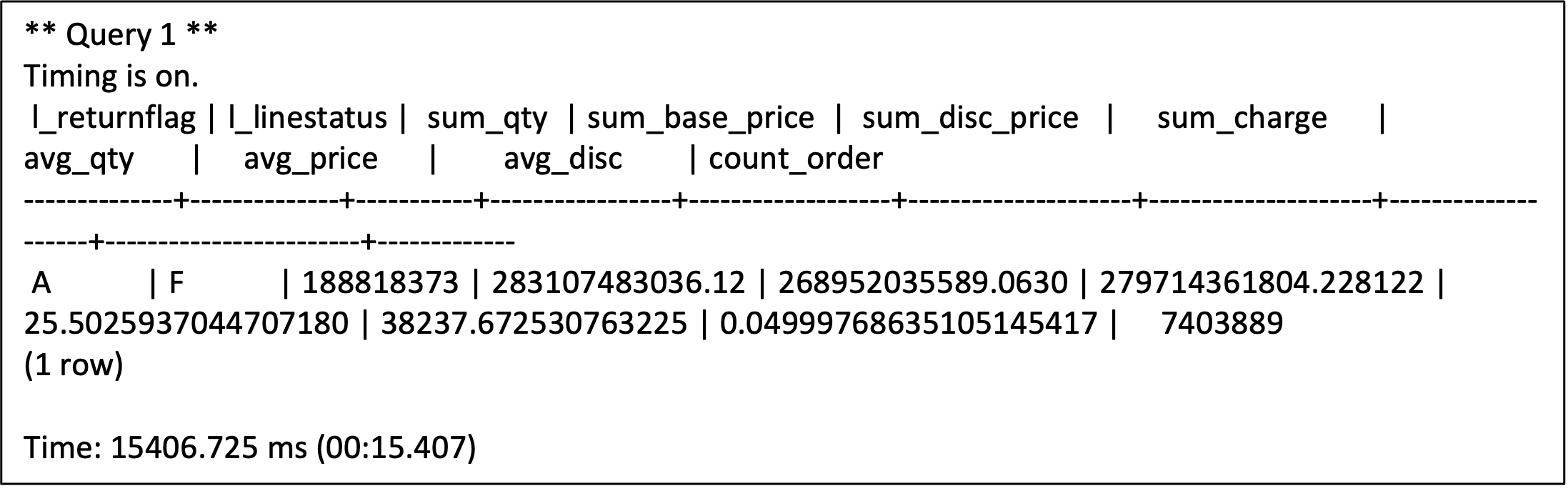}
    \caption{Results of TPC-H Query 1 execution.}
    \label{fig:raw-tpch}
\end{figure}

\systemname supports TPC-H benchmark, a decision support benchmark consisting of a suite of 21 business-oriented ad-hoc queries.
The result file of TPC-H consists of a retrieved table as a response to the query and the corresponding processing time as shown in Figure~\ref{fig:raw-tpch}.
The system also provides visualization of query execution plans obtained using the \verb|EXPLAIN| command.
Currently, the system supports query plans of three high-ranking relational DBMSs: PostgreSQL, MariaDB, and MySQL\cite{db-ranking}.

\subsubsection{TPC-H Result Visualization}
This section introduces the data processing techniques and visualization designs employed for the comparative view on the right side of the interface.
Similar to the data processing approach used in sysbench, \systemname parses the duration from the raw result files and then stores the data in an array format based on query number sequence.

\begin{figure}[t]
    \centering
    \includegraphics[width=\linewidth]{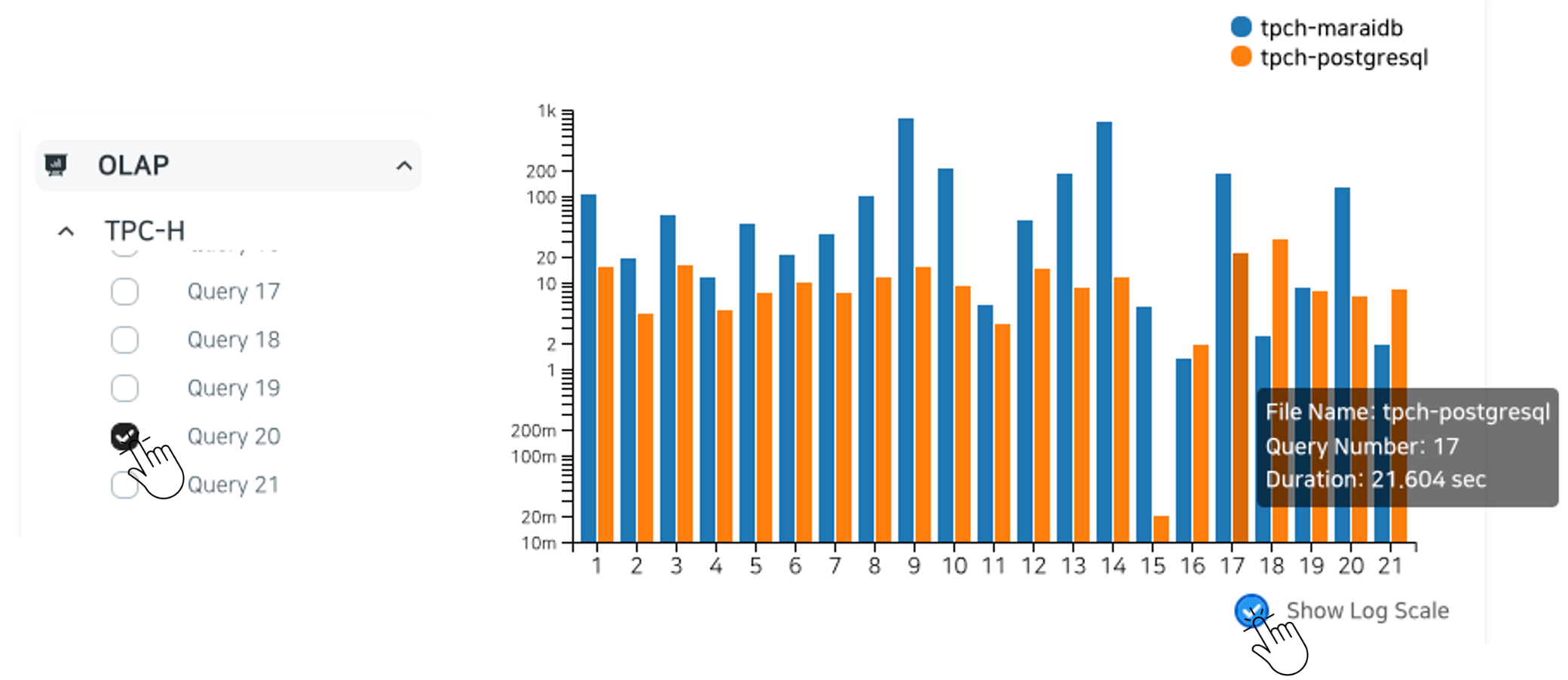}
    \caption{Interaction features of grouped bar chart.}
    \label{fig:tpch-compare}
\end{figure}

The processed data are visualized as a grouped bar chart, providing a broad overview of DBMS performance comparison.
The interactive grouped bar chart is illustrated in Figure~\ref{fig:tpch-compare}.
The chart divides groups by file and maps unique colors for each group to enable easy identification.
We implement a showing log scale feature to address the limitation of representing substantial differences in duration among queries and DBMSs.
Users can turn the feature on or off using the `Show log scale' button to see the chart in the log scale or linear scale respectively.
Additionally, tooltips are implemented, offering detailed information such as file name, query number, and duration.
Users can click bars within the grouped bar chart to select the desired comparison criteria.

\begin{figure}
    \centering
    \subfloat[Duration bar chart]{\includegraphics[width=0.38\linewidth]{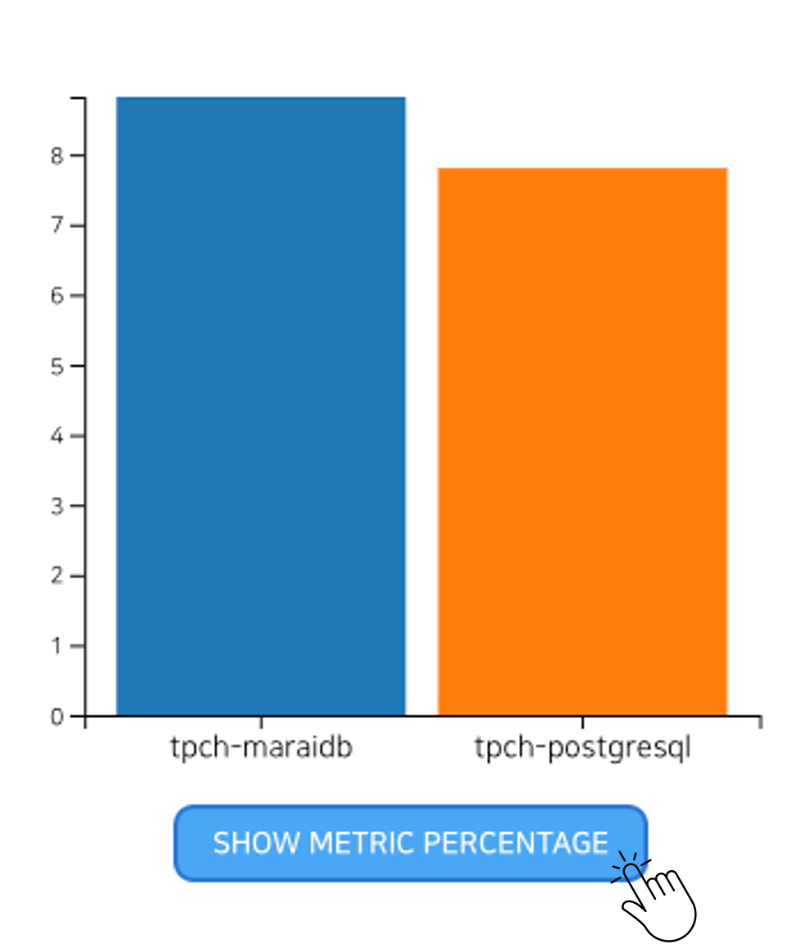}}
    \hskip1ex
    \subfloat[Percentage Distribution]{\includegraphics[width=0.52\linewidth]{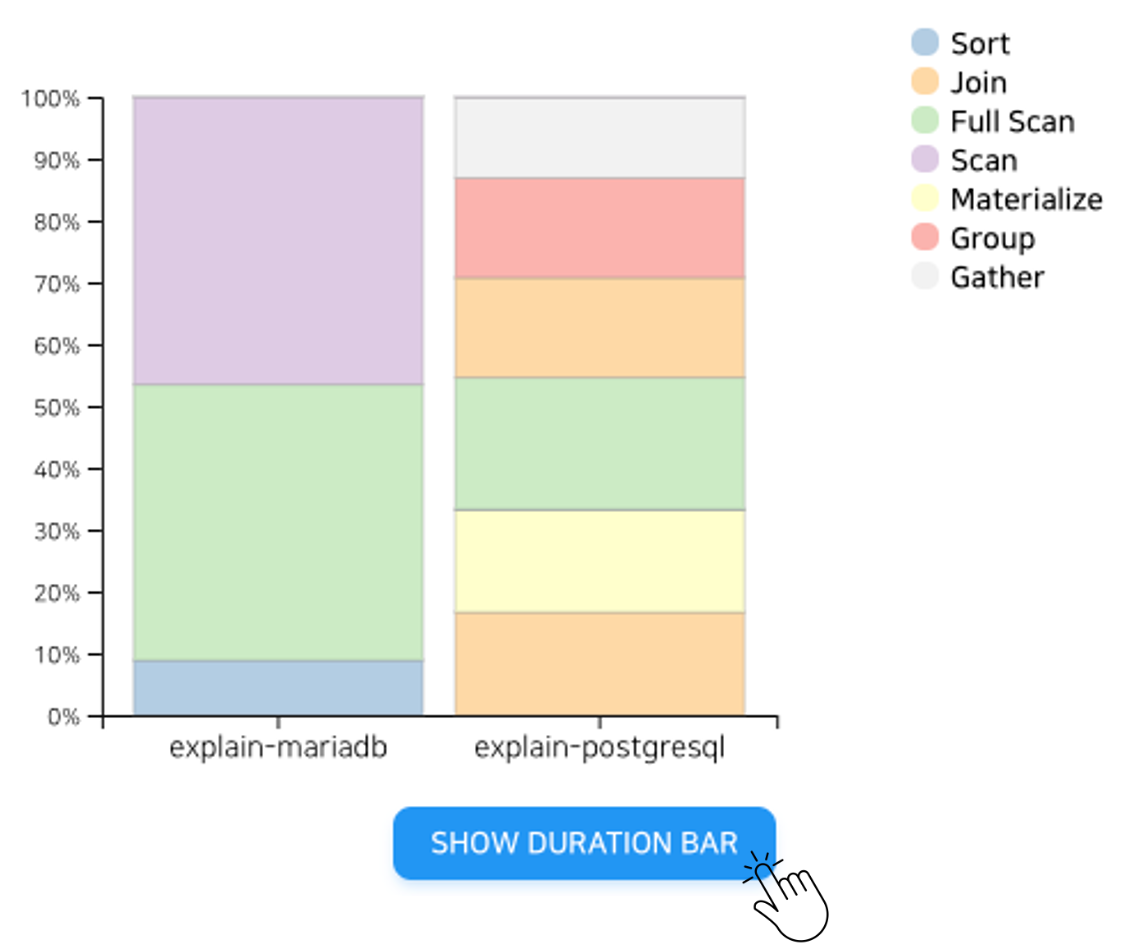}}
    \caption{Bar chart for selected query with bars representing duration and percentage distribution of selected metric.}
    \label{fig:tpch-selected}
\end{figure}

Through the interaction of clicking bars within a grouped bar chart or checkbox in the sidebar, users can perform an in-depth comparative analysis within the selected bar chart positioned at the bottom (Figure~\ref{fig:duration}).
The selected bar chart displays detailed comparisons, enabling users to perform a more comprehensive examination of the selected query.

\subsubsection{Data Processing for \texttt{EXPLAIN} Results}
\begin{figure}
    \centering
    \subfloat[PostgreSQL]{\includegraphics[width=0.45\linewidth]{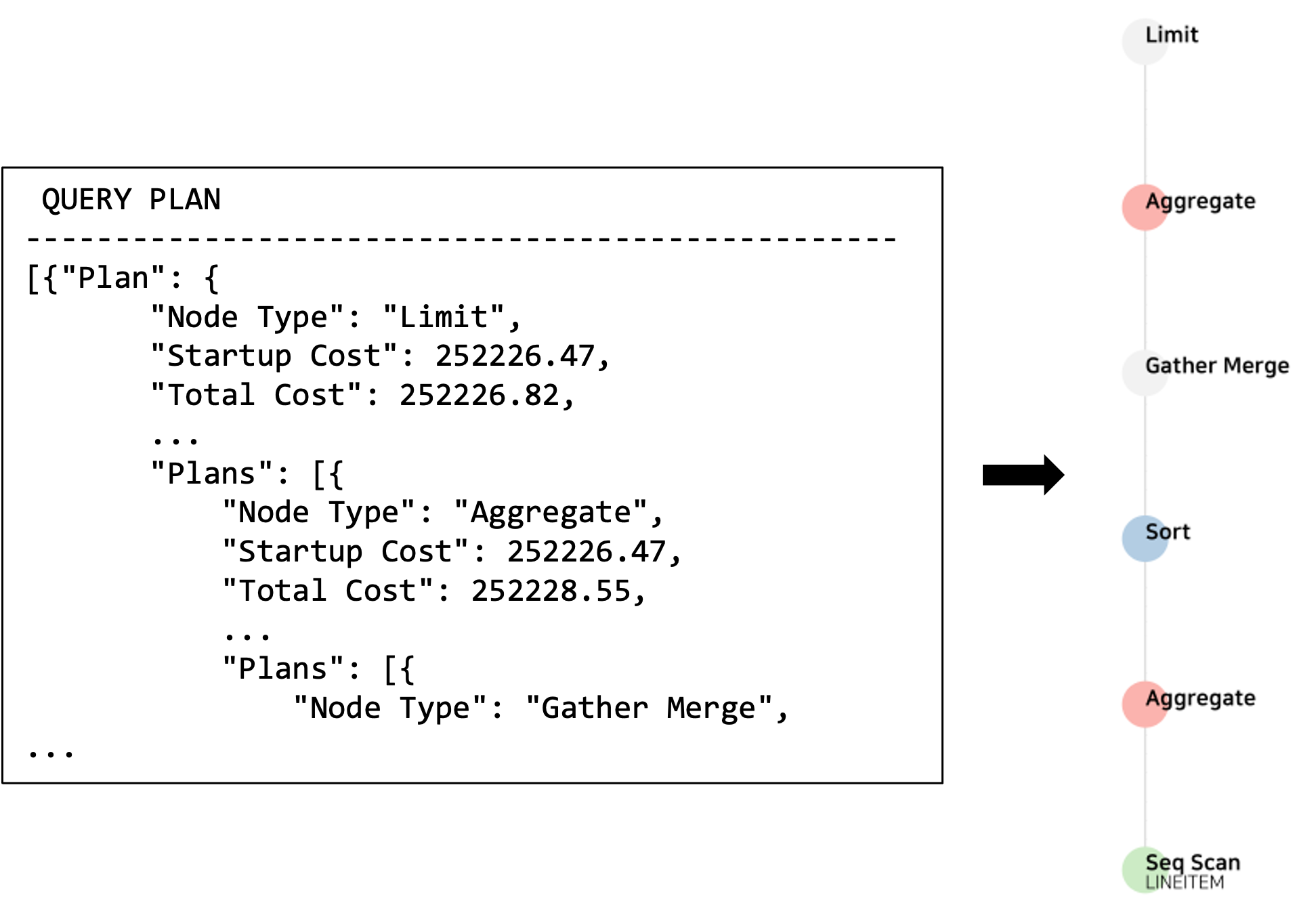}}
    \hskip1ex
    \subfloat[MariaDB]{\includegraphics[width=0.45\linewidth]{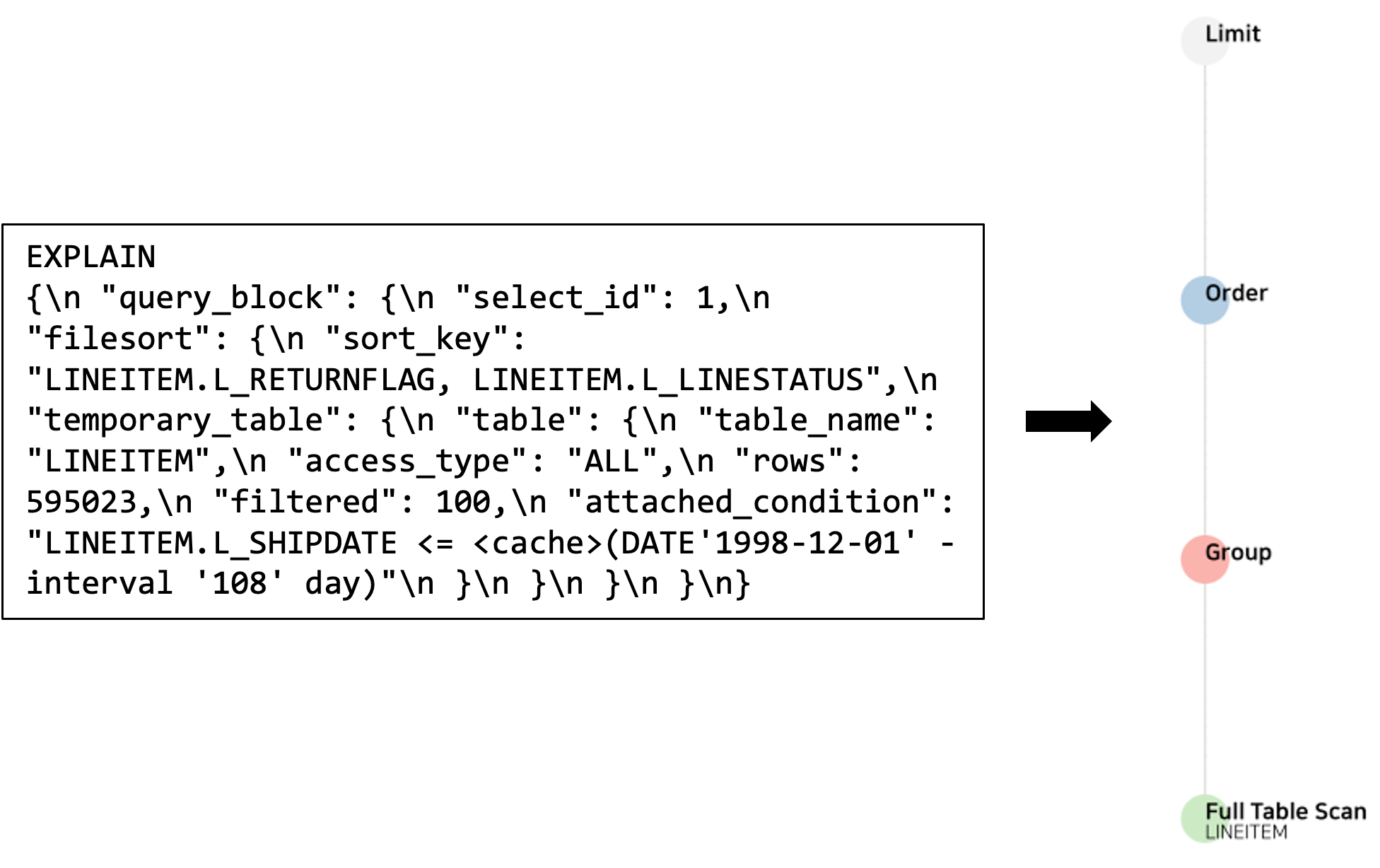}}
    \caption{Query execution plans and generated trees for TPC-H Query 1.}
    \label{fig:raw-qp}
\end{figure}

This section describes the techniques developed for processing the \verb|EXPLAIN| results.
Figure~\ref{fig:raw-qp} illustrates the query plans obtained from PostgreSQL and MariaDB for TPC-H Query 1.
Notably, the challenge lies in addressing the unique format of \verb|EXPLAIN| results and the variation in terminology used for operations, making direct comparisons of query plans complicated.
Furthermore, the original result format does not adhere precisely to JSON standards, preventing the data from directly using the D3 hierarchy or \verb|stratify|.

To overcome these challenges, we first introduce terminology mapping functions that facilitate easier comparison.
The function ensures that even when the terminologies differ for the same operators, the system can identify operators as the same operators.
A more significant contribution lies in developing techniques to parse and rearrange \verb|EXPLAIN| result files for all the DBMSs \systemname supports.
We employ distinct techniques tailored to each unique result format, resulting in a uniform hierarchy format for the query plan data.
This standardized format allows data to be passed to the D3 hierarchy, enabling the construction of root nodes and visualization as a tree.
All additional properties of the operator are stored within the nodes and used for in-depth analysis.

\subsubsection{\texttt{EXPLAIN} Result Visualization Designs}
\begin{figure}[t]
    \centering
    \includegraphics[width=0.9\linewidth]{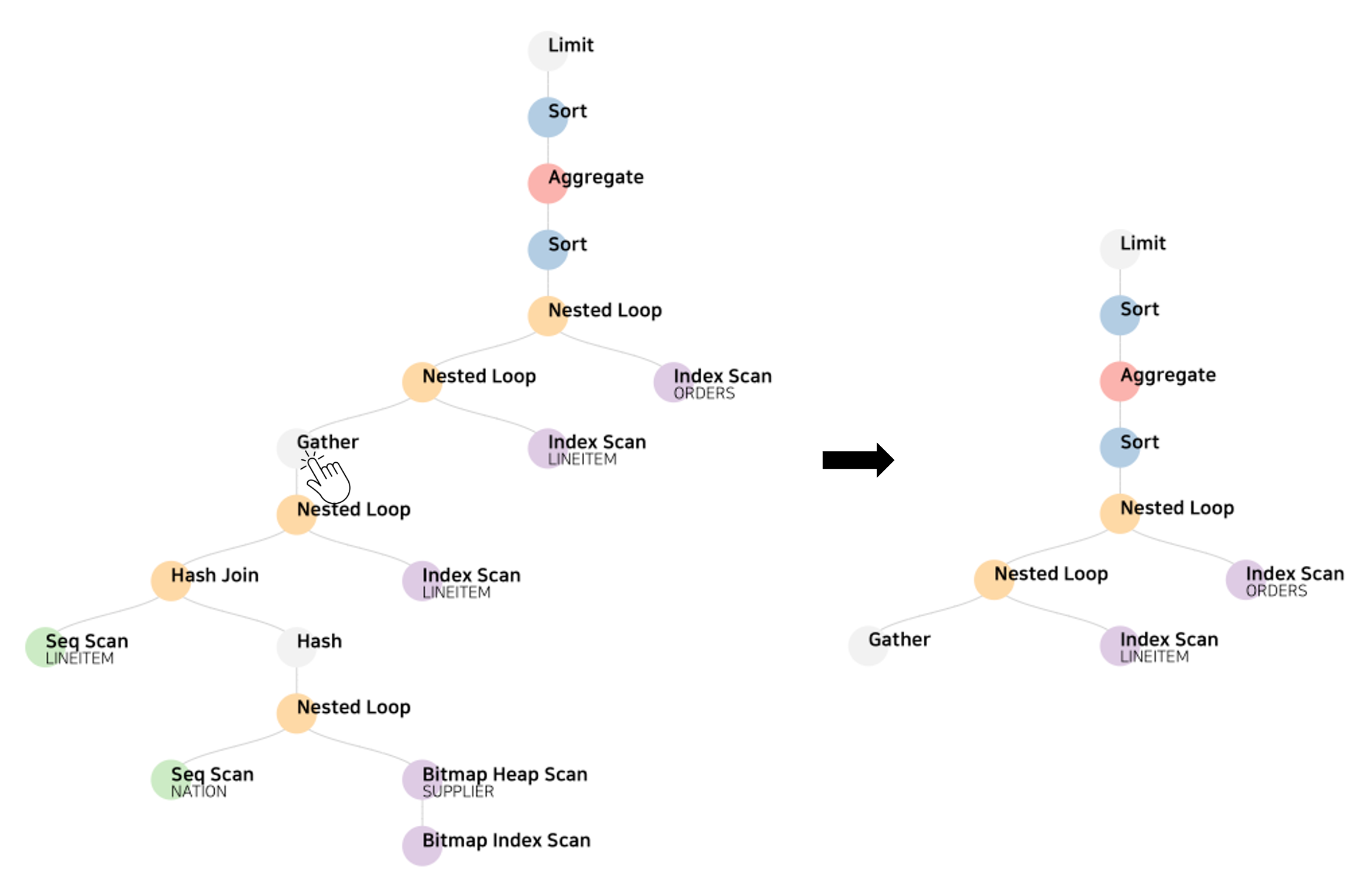}
    \caption{Collapsible tree.}
    \label{fig:tree}
\end{figure}

\begin{figure}[t]
    \centering
    \includegraphics[width=0.9\linewidth]{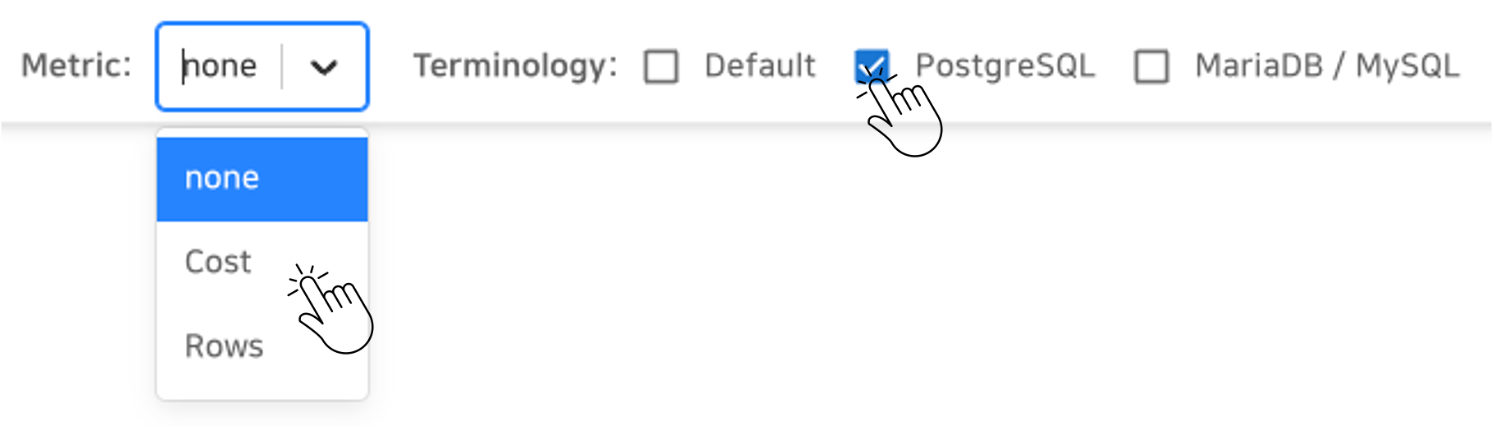}
    \caption{Interaction features of metric and terminology selection.}
    \label{fig:tree-select}
\end{figure}

\begin{figure}[t]
    \centering
    \includegraphics[width=0.9\linewidth]{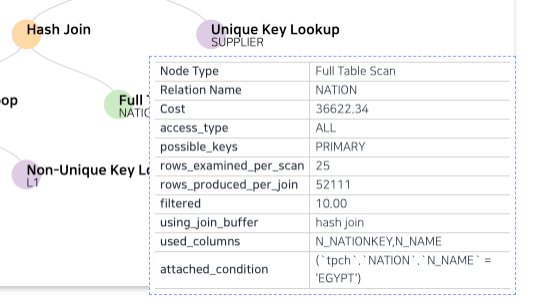}
    \caption{Tooltips with node that contains additional properties of operator.}
    \label{fig:tree-tooltip}
\end{figure}

We describe the query plan visualization designs and user interactions.
The goal is to build an interactive and visually intuitive analysis tool for users to better understand query plans and examine various operations' impact on performance.
The design choice of the tree is motivated by the inherent tree format of query plans, which has a root node and child nodes connected by edges.
Figure~\ref{fig:tree} illustrates the interactive features of the tree, where users can expand or collapse the tree by clicking on nodes.
This enables users to focus on specific segments of the query plan they want to analyze.
The distinct node colors correspond to different operators, enhancing the comparative analysis between various DBMSs.
Each node represents an operator, with the node size proportional to the selected metric.

As shown in Figure~\ref{fig:stacked}, a stacked bar chart appears by clicking on the `Show metric percentage' button.
The chart is designed to show each operator's relative chosen metric (\,  e.g., cost or accessed rows estimated by planner) proportions, assuming a total cost of 100 percent.
In Figure~\ref{fig:tpch-selected}, users can select a metric to visualize.
Furthermore, to reduce the mismatch arising from diverse terminologies, users can standardize the terminology used throughout the tree by selecting the terminology they want to use.
For detailed analysis, the tooltip illustrated in Figure~\ref{fig:tree-tooltip} appears when the user hovers over a specific node.
Tooltips include additional properties such as cost, rows, and attached condition.

\subsection{System Architecture and Implementation}
The architecture of \systemname consists of two primary modules: preprocessing and visualization.
The preprocessing module serves two functions.
First, it extracts meaningful data from the uploaded files.
Second, it parses and reorganizes \verb|EXPLAIN| result files into a hierarchical format.
The visualization module is dedicated to presenting processed data through an interactive interface.
This module is developed as a single application in JavaScript, utilizing D3.js for data-driven visualizations and the React library for building user interfaces.
The system layout is optimized for full-screen viewing, particularly at a resolution of 1920 × 1080.
This ensures that users can easily access and interact with the comprehensive data visualizations provided by the system.

\section{Evaluation}
\label{sec:eval}
To evaluate the efficacy and user-friendliness of our visual analytics tool, we engaged two experts in the field for interviews.
Prior to these interviews, the experts were provided with a concise overview and detailed explanations of the tool's functionalities and the data visualizations implemented.
During the informal interviews, we gathered valuable insights, including suggestions for additional features, opinions on the visual design, and feedback on user interaction with our system.
Both experts concurred that our tool has the potential to significantly enhance analytics for its target users, particularly emphasizing the need for further refinement in user interaction to ensure a smoother, more intuitive experience.

For a more structured assessment, we formulated three key questions, each linked to specific requirements of \systemname:
\begin{itemize}
    \item[R1.] \textbf{Visualize benchmark results.} Does the \systemname effectively visualize the benchmark results?
    \item[R2.] \textbf{Support comparison.} Does the \systemname support easy comparison of benchmark results?
    \item[R3.] \textbf{Provide interactive dashboard.} Are the interactive features of \systemname useful and enhance user experience?
\end{itemize}

\subsection{Fulfillment of Requirements}
We asked participants to respond to a series of statements to determine if \systemname met its intended requirements.
This was done using a 5-point Likert scale survey.
Analysis of the survey results revealed a general consensus among participants that \systemname effectively meets the set requirements.
This was reflected in the average scores: 4.5 for R1, 4 for R2, and a perfect score of 5 for R3.
These scores indicate that there is relatively little variation in the satisfaction levels across different requirements.
Additionally, the participants' responses reflected a favorable opinion of the tool's usability, with an average rating of 4.5.

\subsection{Discussions}
Future iterations of \systemname offer possibilities for refinement and expansion.
One key area of enhancement involves integrating the actual execution times of database operators into the system.
This feature will allow users to more accurately assess the effectiveness of a query optimizer's selected plan, going beyond simple cost estimations.
Furthermore, based on practical suggestions received during our case studies with experts, we plan to refine the query plan visualization aspect of the system.
By enabling the aggregation of costs from child nodes in a collapsed tree view, \systemname will offer a more comprehensive and nuanced analysis of costs, improving both user interaction and analytical depth.

Another significant addition under consideration is the capability to visualize and compare a variety of potential query plans generated by a DBMS optimizer.
This enhancement would shed light on the spectrum of query strategies evaluated by the optimizer, providing users with deeper insights.
Also, these planned improvements and new features are poised to take \systemname to the next level in terms of functionality and user experience.
We are committed to exploring and implementing these ideas in our forthcoming developments of \systemname.

\section{Conclusion}
\label{sec:conclusion}
We presented \systemname, an interactive visual analytics system designed for the comparative analysis of DBMSs performance.
The contribution of this paper is threefold.
First, we developed techniques to extract meaningful data from raw result files and transform it into the required data format for visualizations.
Second, we implement novel and unified visualization designs to explore results obtained from OLAP and OLTP benchmark executions, as well as \verb|EXPLAIN|.
Lastly, we developed a user-friendly application with interactive features to enhance the user experience.
Through case studies conducted with domain experts, we demonstrated the efficacy and usability of \systemname.
The implementation of our system is publicly available on GitHub at github.com/ratdbs/dbenchvis-front.

\bibliographystyle{abbrv-doi}

\bibliography{references}
\end{document}